\documentclass[twocolumn]{aastex631}
\usepackage{natbib}
\usepackage[normalem]{ulem}

\usepackage{booktabs}
\usepackage{longtable}
\usepackage{amsmath, mathtools, amsthm, amssymb}
\usepackage{enumitem}
\usepackage{graphics}
\graphicspath{{./}}

\usepackage{lipsum}
\usepackage{comment}
\usepackage{multirow}
\usepackage{hyperref}
\usepackage{changepage}

\hypersetup{colorlinks=true, allcolors=cyan}

\def\HAGN{\textsc{Horizon-AGN}} 
\def\NH{\textsc{NewHorizon}}

\def\RAMSES{\textsc{RAMSES}}

\definecolor{Green}{rgb}{0.1,0.7,0.2}

\newcommand\edel{\bgroup\markoverwith
{\textcolor{red}{\rule[0.5ex]{2pt}{0.8pt}}}\ULon}

\begin{document}
\title{Formation pathways of the compact stellar systems}

\author[0000-0002-0858-5264]{J. K. Jang}
\affil{Department of Astronomy and Yonsei University Observatory, Yonsei University, Seoul 03722, Korea}

\author[0000-0002-4556-2619]{Sukyoung K. Yi}
\affil{Department of Astronomy and Yonsei University Observatory, Yonsei University, Seoul 03722, Korea}
\email{yi@yonsei.ac.kr}

\author[0000-0002-0041-6490]{Soo-Chang Rey}
\affil {Department of Astronomy and Space Science, Chungnam National University, Daejeon 34134, Korea}

\author[0000-0002-0184-9589]{Jinsu Rhee}
\affil{Department of Astronomy and Yonsei University Observatory, Yonsei University, Seoul 03722, Korea}

\author[0000-0003-0225-6387]{Yohan Dubois}
\affiliation{Institut d’Astrophysique de Paris, Sorbonne Université, CNRS, UMR 7095, 98 bis bd Arago, 75014 Paris, France }

\author[0000-0002-3950-3997]{Taysun Kimm}
\affil{Department of Astronomy and Yonsei University Observatory, Yonsei University, Seoul 03722, Korea}

\author[0000-0003-0695-6735]{Christophe Pichon}
\affiliation{Institut d’Astrophysique de Paris, Sorbonne Université, CNRS, UMR 7095, 98 bis bd Arago, 75014 Paris, France }

\author[0000-0001-6180-0245]{Katarina Kraljic}
\affil{Universit\'e de Strasbourg, CNRS, Observatoire astronomique de Strasbourg, UMR 7550, F-67000 Strasbourg, France}

\author[0000-0003-3474-9047]{Suk Kim}
\affil {Department of Astronomy and Space Science, Chungnam National University, Daejeon 34134, Korea}

\received{XX}
\revised{YY}
\accepted{ZZ}
\submitjournal{ApJ}

\shorttitle{Formation pathways of the compact stellar systems}

\shortauthors{Jang, JK et al.}


\begin{abstract}
The formation pathways of compact stellar systems (CSSs) are still under debate. 
We utilize the \NH\ simulation to investigate the origins of such objects in the field environment. We identified 55 CSS candidates in the simulation whose properties are similar to those of the observed ultra-compact dwarfs and compact ellipticals.
All but two most massive objects (compact elliptical candidates) are a result of a short starburst.
Sixteen are formed by tidal stripping, while the other 39 are intrinsically compact from their birth. 
The stripped objects originate from dwarf-like galaxies with a dark halo, but most of their dark matter is stripped through their orbital motion around a more massive neighbor galaxy.
The 39 intrinsically compact systems are further divided into ``associated'' or ``isolated'' groups, depending on whether they were born near a massive dark halo or not. 
The isolated intrinsic compact objects (7) are born in a dark halo and their stellar properties are older and metal-poor compared to the associated counterparts (32).
The stripped compact objects occupy a distinct region in the age-metallicity plane from the intrinsic compact objects. 
The associated intrinsic compact objects in our sample have never had a dark halo; they are the surviving star clumps of a massive galaxy.
\end{abstract}


\section[]{Introduction}
\label{sec:introduction}
From a few parsec scale (star clusters, usually globular clusters) to ten kilo-parsec scale (massive elliptical galaxies), many astronomical objects with various masses embroider space.
Nearly a century ago, there was a clear separation between star clusters and galaxies in terms of size.
Not only are stellar clusters relatively smaller in mass compared to massive galaxies, but they are also clearly smaller ($R_{\rm eff}\leq 30$ pc) than dwarf spheroidal galaxies of similar mass ($R_{\rm eff}\geq 500$ pc).
However, the gap between these two different types of objects is gradually being filled with observational data points.
Compact elliptical galaxies (cE), which are more massive ($10^{\rm 8}\,\leq\, M_{\rm *}\,\leq\,10^{\rm 10}\,\rm M_{\rm \odot}$) than globular clusters but smaller ($100 \leq R_{\rm eff} \leq 900$ pc) than typical galaxies in the same mass range, are found in galaxy cluster environments \citep{Rood1965,Faber1973,Price2009}.
Additionally, less massive stellar systems ($10^{\rm 6} \leq M_{\rm *} \leq 10^{\rm 8}\ \rm M_{\rm \odot}$) with much smaller sizes (a few 10 pc), known as ultra-compact dwarfs (UCDs), have also been reported \citep{Hilker1999,Drinkwater2000,Phillipps2001,Hacsegan2005}.
When boundaries between the two are gradually fading,
the key question now arises: what defines a galaxy? \citep{Forbes&Kroupa2011,Willman&Strader2012}

Recently, the archive of intermediate mass stellar systems (AIMSS) survey \citep{Norris2014_AIMSS1,Forbes2014_AIMSS2,Janz2016_AIMSS3} attempted to find the possible origins of compact stellar systems (CSSs) within this gap by aggregating vast amounts of observed catalogs with additional observations.
Several possible formation scenarios for these galaxies have been suggested \citep{Bekki2001b,Martinovic2017,Ferre-Mateu2018,Zapata2019,Mahani2021}, but they could be simplified as follows: are they intrinsic or not? 
If CSSs are intrinsically smaller than the main branch of spheroidal galaxies, cEs can be treated as the lower mass end of classical luminous elliptical galaxies\citep{Wirth1984,Kormendy2009,Kormendy2012}, while ultra-compact dwarf galaxies can be considered the high mass end of globular clusters \citep{Marks2012, Norris2019}.
If they are not intrinsically smaller, the most plausible scenario is that they are the tidally stripped remnants of more massive galaxies, as some CSS is often found in tidal streams around neighboring massive host galaxies \citep{Huxor2011b,Paudel2013,Liu2015,Voggel2016,Ferre-Mateu2018}. 
The stellar population of the object, its star formation history, the existence of a supermassive black hole (SMBH), and the surrounding environment can all be used to build possible explanations \citep{Francis2012,Mieske2013,Seth2014, Afanasiev2018,Ferre-Mateu2018,Ferre-Mateu2021,Rey2021,Chen2022}.
However, the main obstacle in revealing the formation scenario in observation is that we cannot trace the entire evolutionary history of the object throughout cosmic time.

To support and clarify the formation and evolution of compact systems, theoretical, more specifically numerical predictions, are attempting to find such objects in simulations.
The idealized simulations in a closed box attempt to elucidate the tidal stripping process of existing dwarf or spiral galaxies with central concentrations \citep{Bekki2001a,Bekki2001b,Pfeffer&Baumgardt2013}, or to address possible formation via star cluster-cluster collisions \citep{Kroupa1998,Fellhauer&Kroupa2002,Bruns2011,Zapata2019}.
However, hydrodynamic simulations on a cosmological scale are mostly inadequate or have failed to provide clues to the given question, mainly due to computational costs.
Just a few years ago, cosmological hydrodynamic simulations were difficult to achieve spatial and mass resolutions sufficient to decompose the detailed structure of compact systems within a satisfactory volume, typical spatial and stellar mass resolutions being 1 kpc and $10^{6-8}\ \rm M_{\odot}$, respectively \citep{Vogelsberger2014,Schaye2015,Dubois2016}. 
Of course, despite this poor condition, some approaches have been attempted to overcome this limitation and address the question \citep{Chabanier2020,Mayes2021}.

We finally appear to have the resolution sufficient to study the formation of relatively-massive CSSs in a hydrodynamic simulation on a cosmological scale.
Using IllustrisTNG \citep{Springel2018} with a spatial resolution of about $\leq 300\, \rm pc$, \cite{Deeley2023} attempted to find the formation pathways of compact elliptical galaxies.
Under the sufficient volume of the simulation, a diversity of formation pathways is found, and the overall distribution of the cEs in the parameter space matches well with observations.

We are attempting to push the analysis to the lower mass end, covering ultra-compact dwarf galaxies, using the \NH\ simulation \citep{Dubois2021}.
By sacrificing the volume coverage of the simulation box, \NH\ increased the spatial and stellar mass resolutions (approximately $\sim34\,\rm pc$ in spatial resolution and $\sim10^{\rm 4}\,\rm M_{\rm \odot}$ in stellar mass for a single particle).
In this paper, we discuss observation data, simulation specifications, and sample selection criteria in Section~\ref{sec:methodology}.
In Section~\ref{section:result}, we present the size-mass relation and stellar metallicity-mass relation, and dynamical mass of the CSSs in \NH\ and discuss on their possible formation scenarios.


\begin{figure*}
\centering
\includegraphics[width=0.8\textwidth]{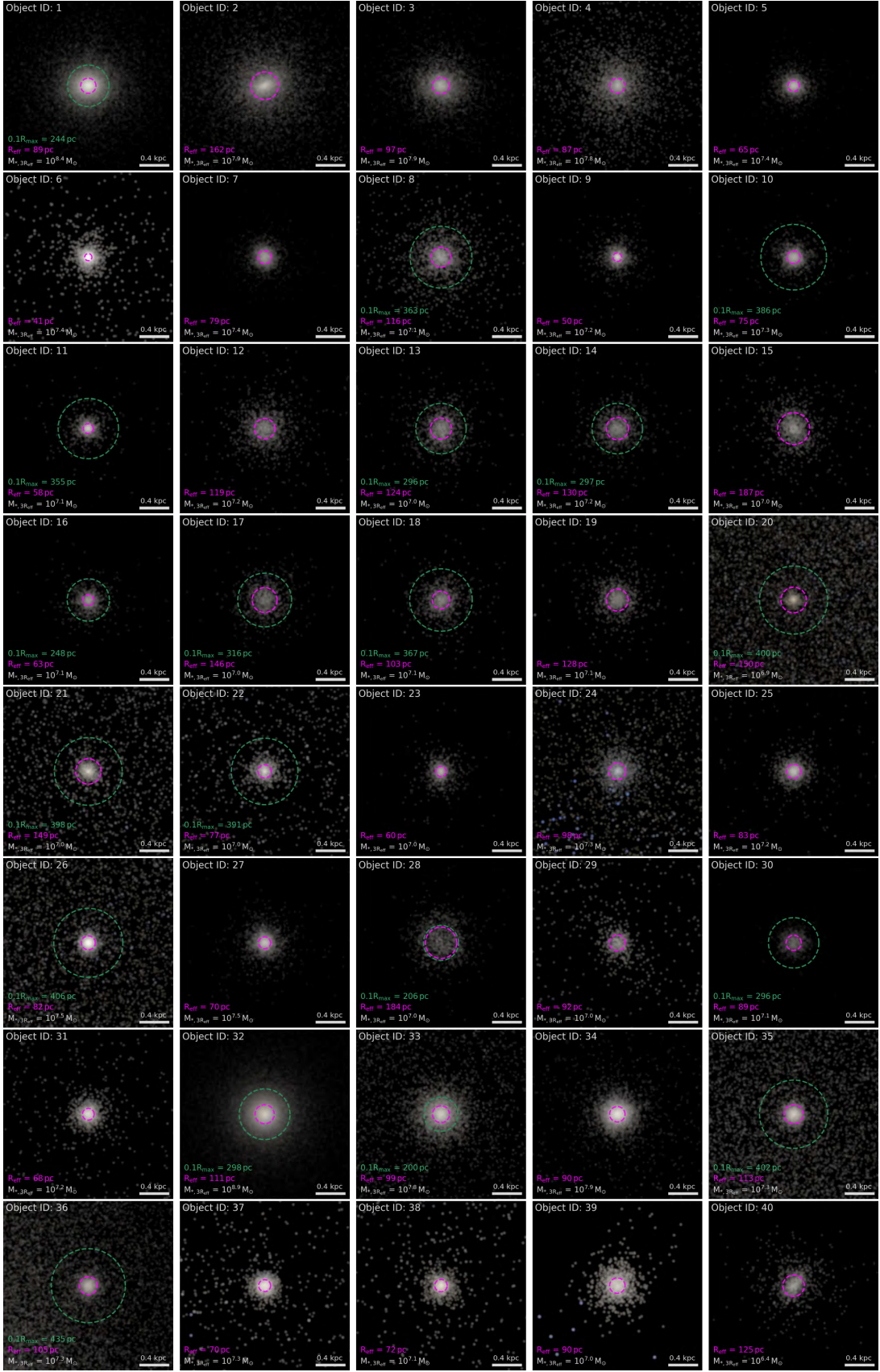}
\caption{
The mock images of 40 random-selected compact stellar systems at the last snapshot ($z=0.17$) in \NH. The samples with two circles are those that are heavily stripped as discussed in the text. The magenta circles show 1\,$R_{\rm eff}$ in the final stage. The green circles show the 0.1\,$R_{\rm max}$ if the sample is formed by tidal stripping, where $R_{\rm max}$ is the half-mass radius when the sample was the most massive in its history. The stellar mass inside 3\,$R_{\rm eff}$ is also given.
}
\label{fig:Figure_all}
\end{figure*}

\section{Methodology}
\label{sec:methodology}

\subsection{Simulation}
\label{subsection:simulation}
\NH\ is a hydrodynamic simulation using \RAMSES\ \citep{Teyssier2002}, an adaptive mesh refinement (AMR) code, covering a spherical region with a radius of approximately $10\, \rm Mpc$. From the initial conditions of the \HAGN\ simulation \citep[][]{Dubois2016}, \NH\ targets a ``field" region as a zoom-in region. The simulation utilized cosmological parameters in accordance with the WMAP-7 data \citep{Komatsu2011}, including a Hubble constant of $\rm H_{\rm 0} = 70.4\,{\rm km\,s}^{-1}\,{\rm Mpc}^{-1}$, a total mass density of $\Omega_{\rm m} = 0.272$, a total baryon density of $\Omega_{\rm b} = 0.0455$, a dark energy density of $\Omega_{\rm \Lambda} = 0.728$, a power spectrum amplitude of ${\sigma}_{\rm 8} = 0.809$, and a spectral index of $n_{\rm s} = 0.967$. The (highest) mass resolutions of the dark matter and stellar particle are $1.2\times10^{6}$ and $1.3\times10^{4}\, \rm M_{\odot}$ for each, respectively. Also, the corresponding maximum spatial resolution for the mesh structure is $34\, \rm pc$ at $z=0$. 
\NH\ has a high time output cadence of 15\,Myr on average. As we will discuss more in the following sections, it allows us to trace the orbital motion of our target objects in great detail.
Star formation (SF) can take place when the hydrogen number density of the gaseous cell is above $10\, \rm H\, \rm cm^{-3}$. SF is modeled using the Schmidt law, with varying star formation efficiency depending on the turbulent Mach number of the cloud and the virial parameter. 
For stellar feedback, the feedback from type-II supernovae (SNe) is included based on the mechanical feedback scheme \citep{Kimm2015}.
A detailed description of \NH\ can be found in \citet[][]{Dubois2021}.
The \NH\ simulation has run to $z=0.17$.

\subsection{Sample Selection}
\label{subsection:sampling}
For the stellar particles, we performed galaxy detection using the AdaptaHOP algorithm \citep{Aubert2004} with the most massive sub-node mode \citep{Tweed2009}. We only considered stellar systems with more than 50 particles.

We first selected the main sample at redshift 0.17, the latest snapshot of the simulation, with the following criteria:
\begin{itemize}
\item $M_{\rm *} > 10^{6}\, \rm M_{\odot}$ and $R_{\rm eff} < 900\,{\rm pc}$, where $R_{\rm eff}$ is the half-mass radius.
\item The mean stellar age of the compact system is larger than 1\,Gyr to avoid transient star clumps. When we inspect all the compact stellar systems detected, 71\% of them are kinematically dispersed within 1\,Gyr. Subsequently, 70\% of those that survived the first Gyr remain compact for the next 2\,Gyr. A choice of a larger value than 1\,Gyr as our selection criterion would lead to a small sample size, but the main conclusions do not change.
\item We use objects that are 100\% free from low-resolution dark matter (DM) particles, including neighboring massive dark halos, if any, (see Section~\ref{subsection:observation} for our classification for ``associated'' and ``isolated'' objects).  
This is to exclude objects possibly formed by numerical artifacts. 
Due to the zoom-in nature of the simulations, low-resolution DM particles can easily travel through the boundary and affect the stability of the encountered gas cells. 
To address this, we define contaminated objects by measuring the maximum contamination fraction from their birth to the last snapshot. 
Nearly 80\% of CSSs initially found in \NH\ are contaminated by low-resolution DM particles.
\end{itemize}

Under these sample selection criteria, we select 55 objects as a reliable sample. 
We were able to follow all of them from their birth to the final snapshot.
We also measure the stellar metallicity and the dynamical mass $M_{\rm dyn}$ inside $3 R_{\rm eff}$, where $M_{\rm dyn} = M_{\rm *} + M_{\rm DM} + M_{\rm gas}$.
Because DM particles are much more massive ($\sim 10^6\ \rm M_\odot$) than star particles in \NH, compact objects in our investigation often have few DM particles inside 3$R_{\rm eff}$. 
Therefore we measure the dark matter mass of our objects inside 3$R_{\rm eff}$ not directly counting DM particles but by fitting the DM profile using the NFW profile \citep{NFW1996}.
Table 1 presents the key properties of the compact objects in \NH.

\begin{figure*}
\centering
\includegraphics[width=0.95\textwidth]{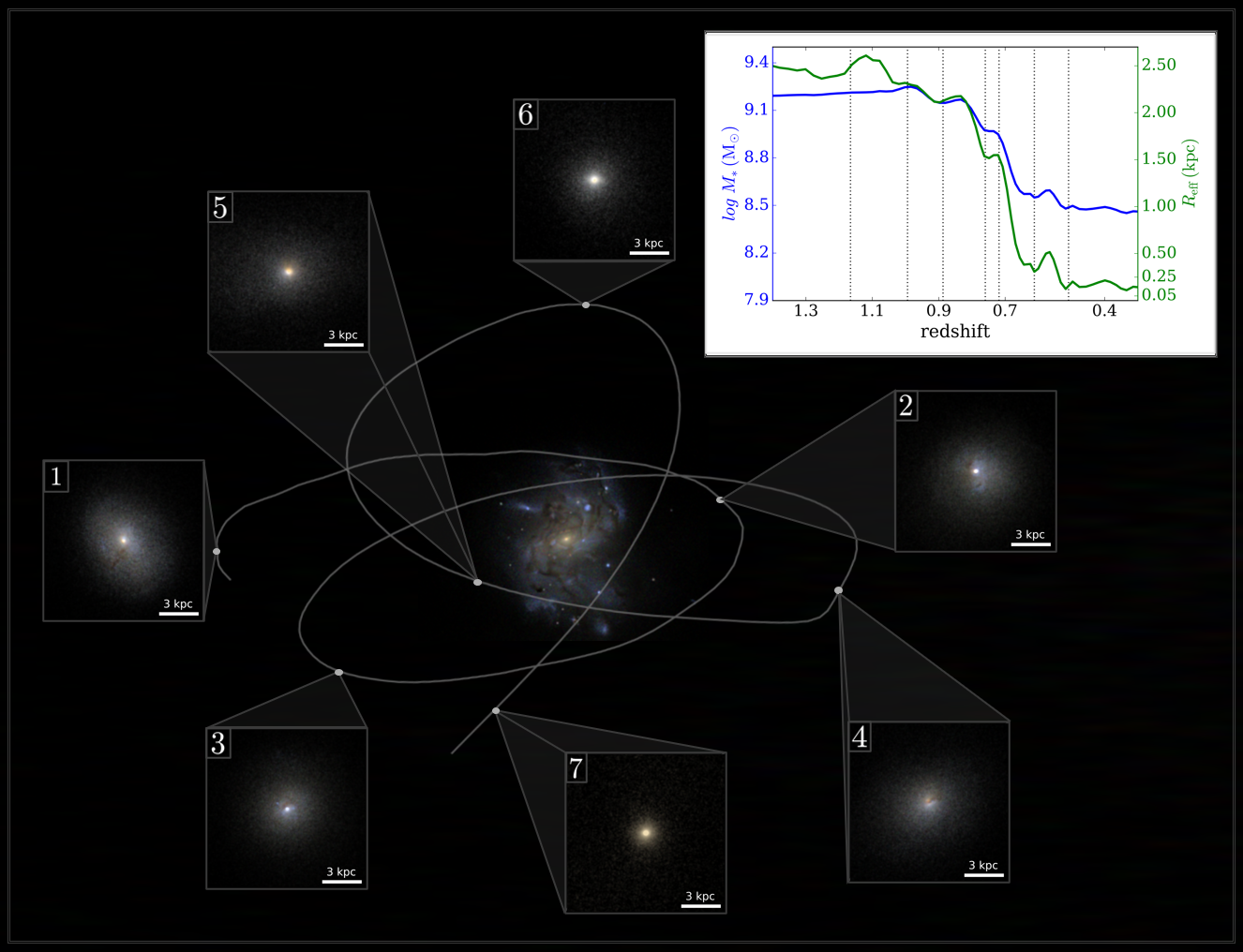}
\caption{
The mock images of a tidally stripped sample in \NH.
Sub-panels 1 through 7 show the time evolution of the compact system orbiting around the massive ($M_{\rm *} \sim 10^{\rm 10.7}\, \rm M_{\rm \odot}$) spiral galaxy. 
The stellar mass (blue solid line) and the half-mass radius (green dashed line) as a function of time are presented in the inset diagram. The seven vertical lines mark the seven snapshots in the main diagram. 
} 
\label{fig:evol_track_stripped}
\end{figure*}
\begin{figure*}
\centering
\includegraphics[width=0.95\textwidth]{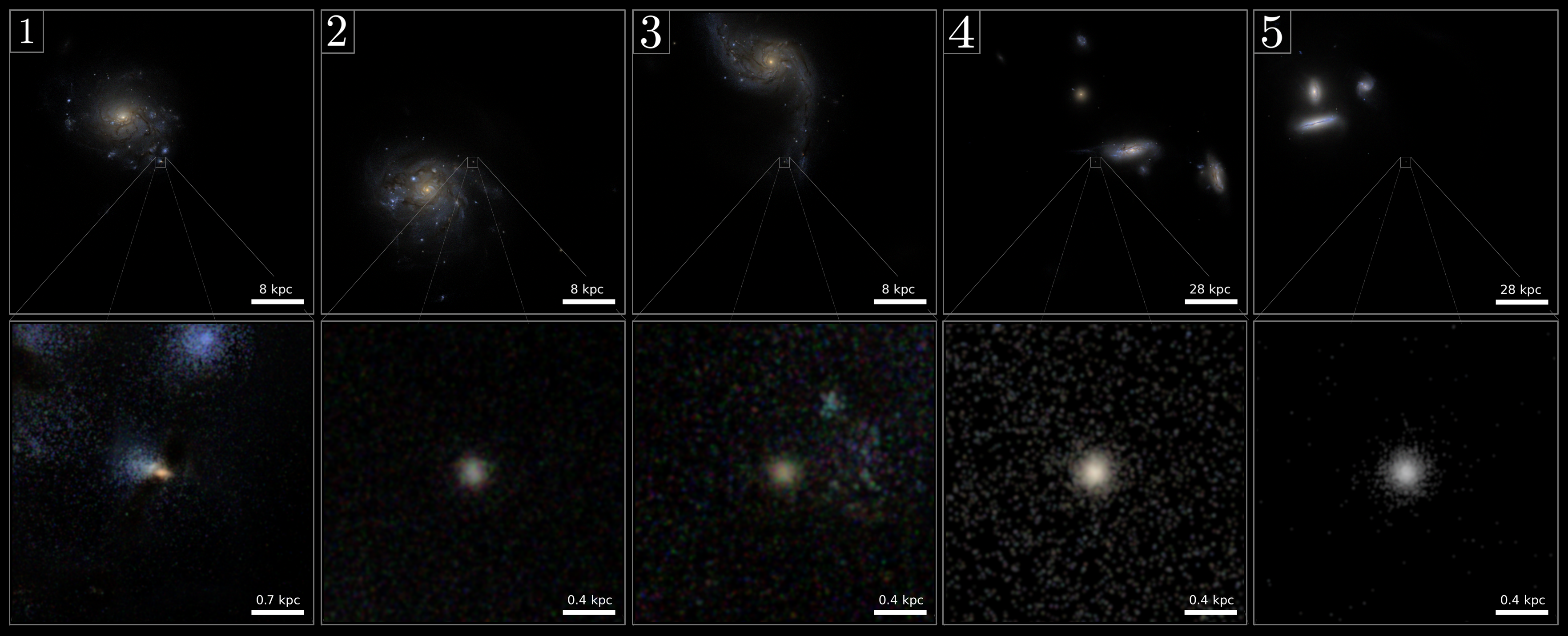}
\caption{
The mock images of an ``intrinsic'' sample in \NH. See the text for definitions. 
Sub-panels 1 through 5 show the time evolution of the compact system.
}
\label{fig:evol_track_associated}
\end{figure*}

\subsection{Observation}
\label{subsection:observation}
For the comparison, we used the observational data covering the wide range of sizes and masses from globular clusters (GCs) to dwarf elliptical galaxies (dEs).
We use the AIMSS catalog Paper-III \citep{Janz2016_AIMSS3} for reference. 
The metallicity is converted from both iron and alpha element abundances: [Z/H] = [Fe/H] + 0.94[$\alpha$/Fe] following previous studies \citep{Thomas2003}.
The sources of the observed data are given in \citep{Janz2016_AIMSS3}. 


\begin{figure*}[t]
\centering
\includegraphics[width=0.9\textwidth]{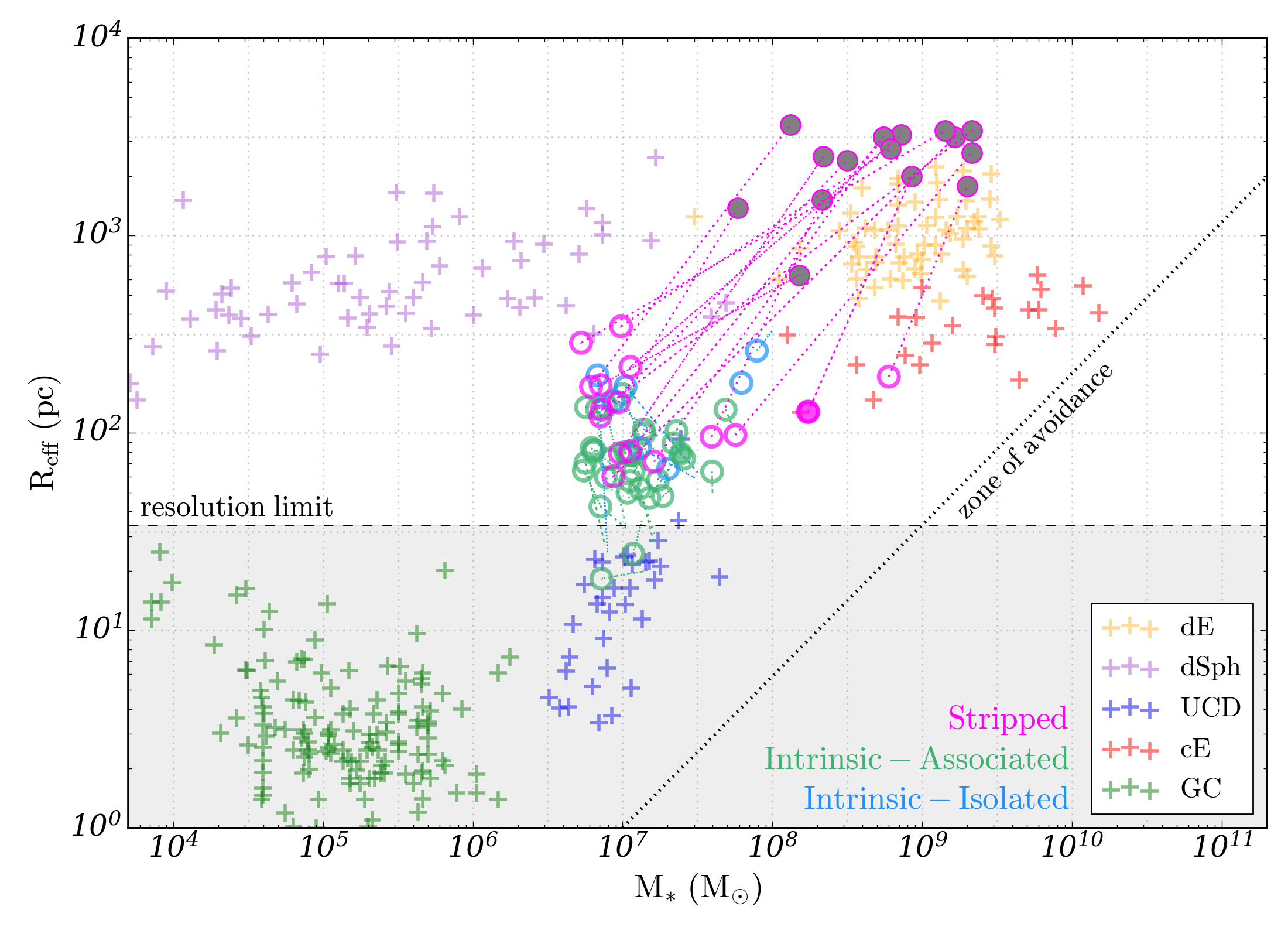}
\caption{
The size-mass relation of the galaxies at $z=0.17$.
The crosses with different colors show the observed catalog data of  \citet[][]{Janz2016_AIMSS3}, and the circles show the simulation samples.
The magenta circles correspond to the stripped objects, the green circles show the intrinsic-associated samples, and the blue circles show the intrinsic-isolated.
The grey-filled circles are the points when the stripped samples were at their maximal points in terms of size ($R_{\rm eff}$), and they are connected with a dotted line to their $z=0.17$ counterparts (open circles).
The intrinsic compact objects (green and blue circles) also have a short dotted line that shows the change of their size and mass since their birth. Their birth properties are not shown here for the sake of clarity.
The magenta- and green-filled circles are the objects shown in Figures~\ref{fig:evol_track_stripped} and \ref{fig:evol_track_associated}, respectively.
The maximum spatial resolution of the simulation (black dashed line) and the zone of avoidance (black dotted line) are provided as a reference.}

\label{fig:size_mass_relation}
\end{figure*}

\section{Results and Discussion}
\label{section:result}

\subsection{Formation mechanisms}
\label{subsection:formation_mechnisms}

We first divide our sample into two sub-samples by comparing the mass and size at their birth and the last snapshot ($z \sim 0.17$).
The first is the sample that goes through tidal stripping. 
We traced all the evolution pathways of the compact systems from their birth and defined the maximal point of their size ({\em stellar} half-mass radius).
If the size and mass decreased by at least half of the maxima, we define them as a tidally stripped sample.
We call this sub-sample ``stripped''.
The second is the sample that is intrinsically small in terms of its size and mass. 
We call this sub-sample ``intrinsic''.
In \NH, 29\% of the sample (16/55) is classified as stripped and the rest (71\%, 39/55) is intrinsic.
Figure~\ref{fig:Figure_all} shows 40 randomly-chosen compact stellar systems in \NH\ at the last snapshot as described Section~\ref{subsection:sampling}.
The magenta circle indicates 1\,$R_{\rm eff}$ of each sample, and for the stripped sample the green circle shows 0.1\,$R_{\rm eff}$ at its maximum size.
We further divide the intrinsic sample into ``associated'' or ``isolated'' depending on whether the object in question was inside the virial radius of any dark halo of mass $\gtrsim 10^{\rm 9}\, \rm M_{\rm \odot}$ at its birth.

We present the mock images of two representative cases of stripped and intrinsic samples in Figures~\ref{fig:evol_track_stripped} and \ref{fig:evol_track_associated}. 
The mock images were made by using SKIRT \citep{Camps2020_SKIRT9} for the SDSS $gri$ bands.
For dust extinction, we used a dust-to-metal ratio of 0.3 with the THEMIS model \citep{Jones2017}.
Figure~\ref{fig:evol_track_stripped} shows the orbital motion of a stripped sample with an apparent central concentration.
From panel 1 to 7, we can see that the dwarf galaxy orbits around the massive spiral galaxy ($M_{\rm *} \sim 10^{\rm 10.7}\, \rm M_{\rm \odot}$) at the center and gradually gets its envelope stripped. 
One can clearly see a change in morphology from a diffuse to a compact system.
The inset diagram shows the time sequence of mass and size of the dwarf galaxy including the seven snapshots shown in the main diagram.
It should be noted that the detailed time sequence can be inspected thanks to the high time cadence of outputs (15\,Myr) of \NH.

We also present an example of the intrinsic sample formed near a host galaxy in Figure~\ref{fig:evol_track_associated}.
It first formed in the giant molecular cloud on the outskirt of a spiral arm at $d_{\rm host} = 15$\,kpc, where $d_{\rm host}$ means the distance to the center of the host galaxy.
Before the minor merger happens, it orbits around the host a few times with negligible change of $d_{\rm host}$ ($d_{\rm host} \sim 20\, \rm kpc$).
The first minor merger happened (between panel 2 and 3), and the fluctuation of the gravitational potential leads the stellar system to migrate away from the host galaxy ($d_{\rm host} \sim 50\, \rm kpc$). 
When two subsequent minor mergers occur (panel 4 and 5), the CSS moves away from the galaxy and its distance from the host galaxy increases to $\sim80\ \rm kpc$. 
The change in size and mass of such objects is more explicitly discussed in the next section. 

\begin{figure*}[t]
\centering
\includegraphics[width=0.9\textwidth]{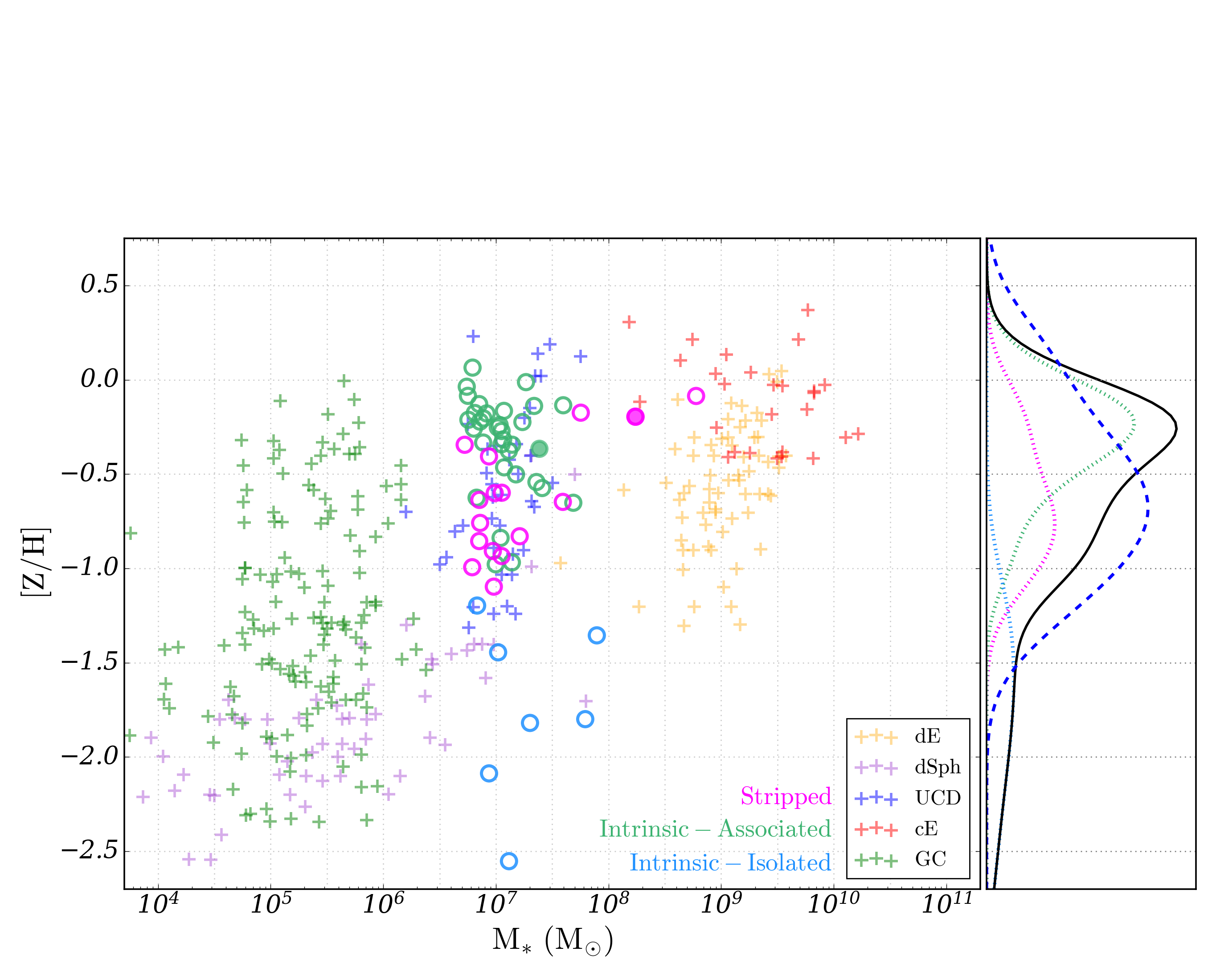}
\caption{
The stellar mass-metallicity relation of the compact objects. 
The marker for each sample is identical to that of Figure~\ref{fig:size_mass_relation}.
The right panel shows $[\rm{Z/H}]$ histograms of the UCD candidates in \NH\ (magenta, green, and blue curves for stripped, intrinsic-associated, and intrinsic-isolated samples, respectively) scaled by their counts (12, 19, and 7). Their sum (solid black curve) is compared with the observed UCDs (blue dashed curve) for the same PDF weight.
}
\label{fig:metal_mass_relation}
\end{figure*}

\subsection{Size-mass plane}
\label{subsection:size_mass}
The size-mass distribution of the compact samples is shown in Figure~\ref{fig:size_mass_relation}, where we present the compact systems in \NH\ compared with the observational data.
The magenta (at $M_* \sim 10^{8.5}\, \rm M_\odot$ and $R_{\rm eff} \sim $ 100 pc) and green filled circles (at $M_* \sim 2.5 \times 10^7 \rm M_\odot$ and $R_{\rm eff} \sim 25$ pc) correspond to the stripped and intrinsic samples presented in Figures~\ref{fig:evol_track_stripped} and \ref{fig:evol_track_associated}.

The open circles show the compact systems in \NH\ at the final snapshot ($z=0.17$).
The magenta, green, and blue circles represent the stripped, intrinsic-associated, and intrinsic-isolated samples.
The stripped samples are connected with dotted lines to dark-filled circles that show their properties of the time when they had the maximum size (that is, in the past).
The horizontal black dashed line corresponds to the maximum resolution that the \NH\ simulation can reach (34 comoving pc). 
We also present the ``zone of avoidance'' in terms of dynamical stability proposed by \citet{Burstein1997}. 
We find that the two massive samples (above $10^8\ \rm M_{\rm \odot}$) have a size and mass comparable with the observed compact elliptical galaxies. 
The less massive samples are comparable to the observed UCDs in terms of mass but show a little offset from the observed distribution in size, probably because of the limited spatial resolution of the simulation.  

The intrinsic samples are born roughly with the present size and mass.
On the other hand, the stripped samples are born like ordinary dwarf galaxies but become compact through tidal stripping.
The time spent between the maximum-size epoch (dark circles) and the epoch of the present size (open circles) is $0.9 \pm 0.7$ Gyr which is roughly a crossing time in the host-satellite system we have from \NH. 
This means that the transformation of a dwarf from an extended to a compact system is dramatic and quick if conditions are met.
This requires a further remark. The two most massive CSSs in our sample including the one in Figure~\ref{fig:evol_track_stripped} are the cases where the central part survived to become a compact object after stripping. The other stripped objects are instead the surviving star clumps that were off-center in the dwarf galaxies.

\subsection{Mass-metallicity plane}
\label{subsection:metal_mass}
We also check the stellar mass-metallicity distribution of the compact systems in Figure~\ref{fig:metal_mass_relation}.
The definition and color scheme for the circles and crosses are the same as those of Figure~\ref{fig:size_mass_relation}.
We consider the solar metallicity value to be 0.02, following \cite{Anders&Grevesse1989}.
We find that the \NH\ sample of compact systems shows a wide variety of metallicity, from $[\rm{Z/H}] = -2$ to 0.5, covering the observed values of UCDs (blue crosses) with similar masses.
The two most massive stripped objects (see Section~\ref{subsection:sampling} for definition) have a stellar mass that is more comparable to those of compact ellipticals.
We present the 1-D histogram of the 53 UCD candidates (excluding the two cE candidates) using kernel density estimation with Silverman's rule of thumb in the right section of the figure.
The sum of the three groups (solid black curve) has a comparable shape of PDF to that of the observed UCDs.

The intrinsic-associated objects (N=32) have high metallicities that are typical of massive galaxies instead of dwarf galaxies.
The intrinsic-isolated objects (N=7) have low metallicities of $[\rm{Z/H}] <-1.0$ which is more comparable to those of dwarf galaxies.
As we will discuss in Section~\ref{subsection:dynamic_mass}, some of them have only a negligible dark halo in the final snapshot ($z=0.17$).

\begin{figure}
\includegraphics[width=0.45\textwidth]{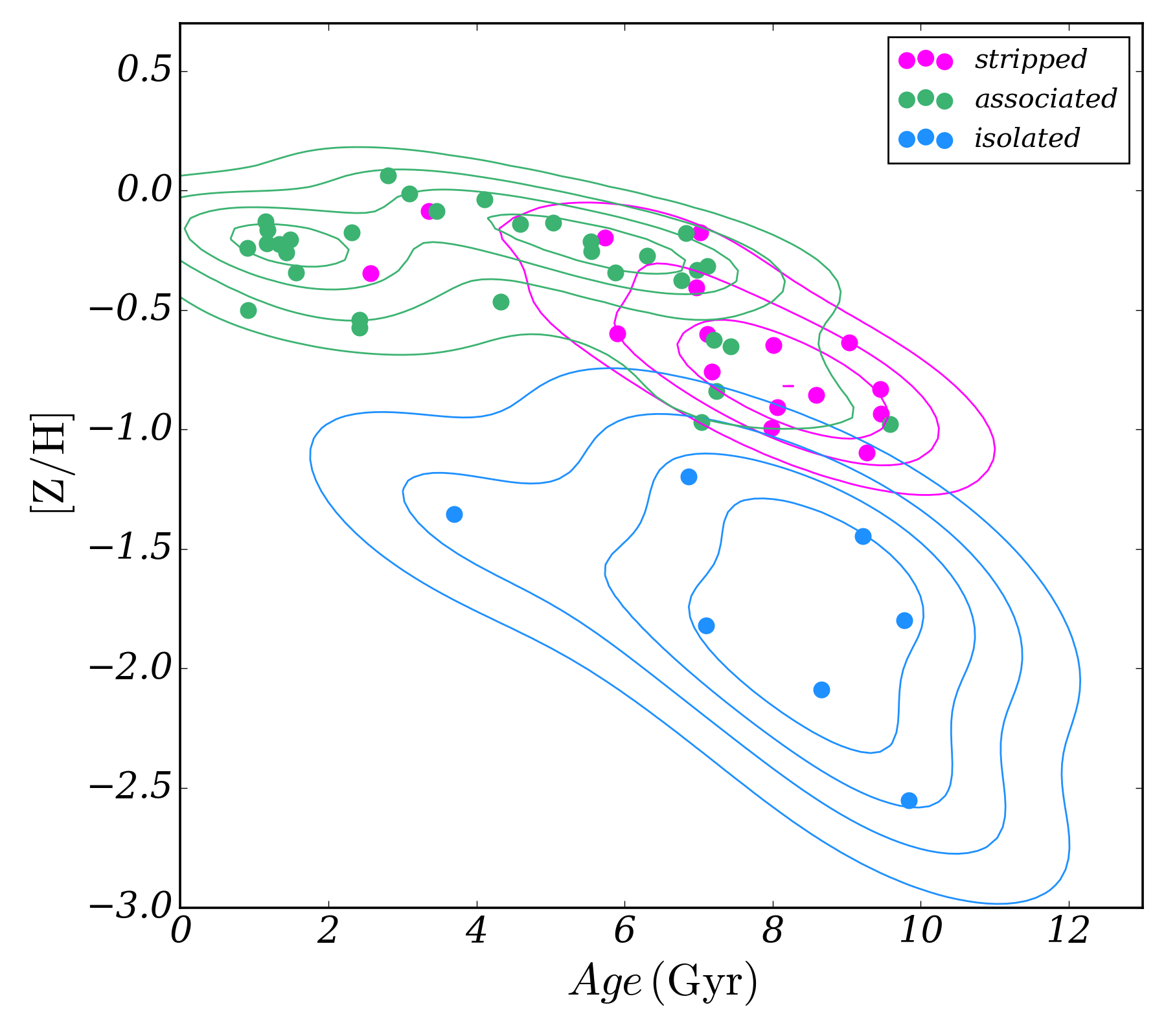}
\caption{
The metallicity and age distribution of 38 UCD candidates.
The magenta, green, and blue colors represent the ``stripped'', ``intrinsic-associated'', and ``intrinsic-isolated'' samples.
The contours for each color are the kernel density estimation of each sample.
}
\label{fig:metal_age}
\end{figure}

We also present our sample in the age-metallicity plane in Figure~\ref{fig:metal_age}.
Both age and metallicity are mass-weighted mean values within $3\,R_{\rm eff}$.
Note that the three groups are reasonably distinct in this diagram, too. 
The mean ages and metallicities (with 16\% and 84\% percentiles) are $4.2^{+2.9}_{\rm -2.9}$\,Gyr and $[\rm{Z/H}]=-0.3^{+0.1}_{\rm -0.3}$ for the intrinsic-associated samples (green symbols/contours), $8.7^{+1.1}_{\rm -1.9}$\,Gyr and $[\rm{Z/H}]=-1.8^{+0.5}_{\rm -0.3}$ for the intrinsic-isolated sample (blue symbols/contours), and $8.0^{+1.3}_{\rm -1.0}$\,Gyr and $[\rm{Z/H}]=-0.7^{+0.3}_{\rm -0.2}$ for the stripped samples (magenta symbols/contours), respectively. 

\subsection{Dynamical mass ratio}
\label{subsection:dynamic_mass}
\begin{figure}
\includegraphics[width=0.45\textwidth]{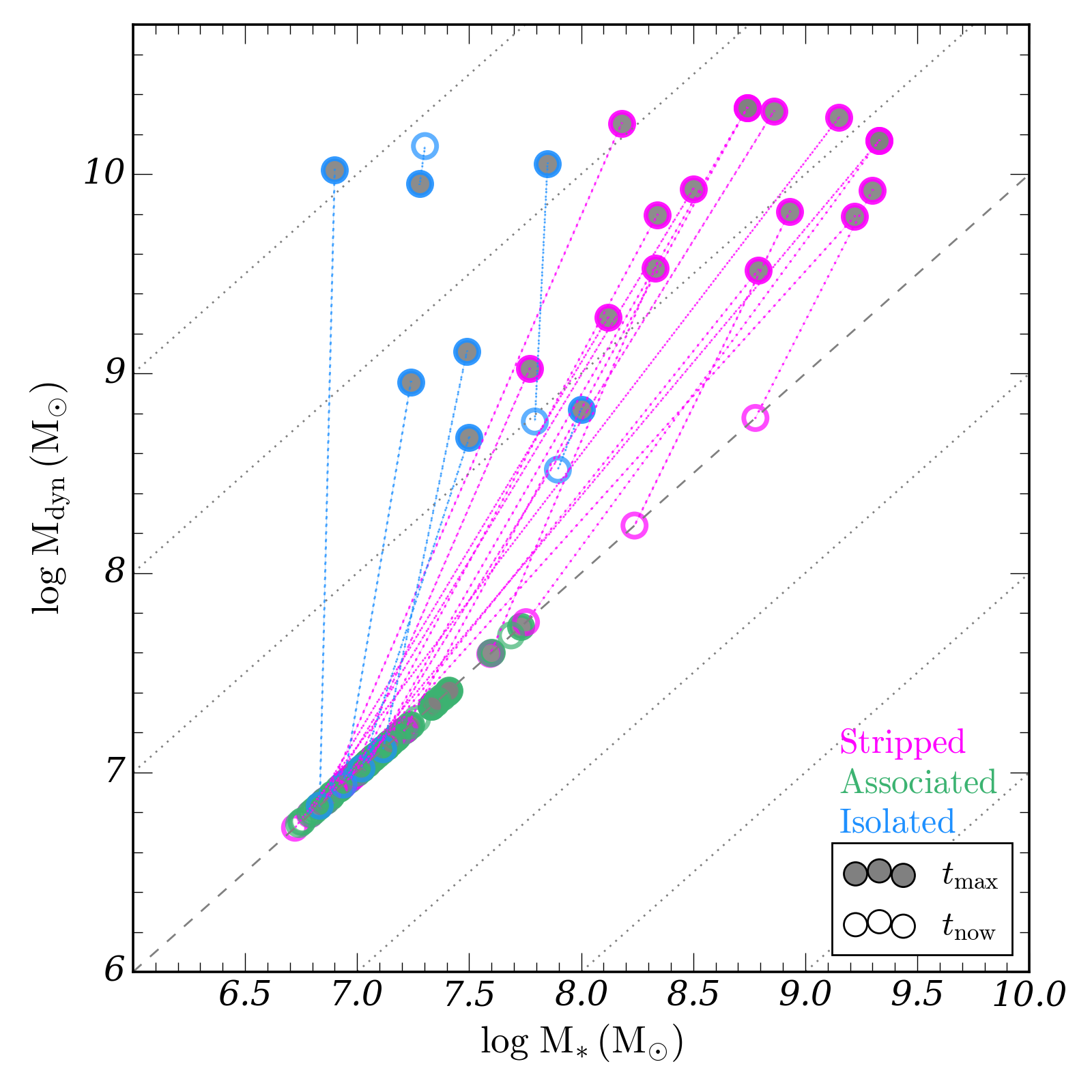}
\caption{
The ratio between the stellar and the dynamical masses inside $3 R_{\rm eff}$.
The symbols follow the same scheme as in previous figures.
The open circles show the present properties. 
The closed circles show the properties of the time when they were maximum in size in the past in the case of the stripped objects and the birth properties in the case of the intrinsic objects.
}
\label{fig:dynamical_vs_stellar_mass}
\end{figure}

We show our samples in the dynamical vs. stellar mass plane in Figure~\ref{fig:dynamical_vs_stellar_mass}.
For the stripped samples, the open circles show the final stage ($z=0.17$), and the filled circles show the values at their maximum size as in Figure~\ref{fig:size_mass_relation}.
They had $M_{\rm dyn}/M_{\rm *}\sim 10$ when they had their maximum size but eventually lost most of their dark matter content through tidal stripping. 
This is not just from the size decrement of the stellar system (in $R_{\rm eff}$) but more importantly from the stripping of dark matter particles of the system itself.
This is consistent with the earlier study of \citet{Smith2016} in the sense that DM halo stripping precedes stellar stripping when a small galaxy orbits around its host galaxy.
Obviously, all of our stripped samples have a massive neighboring galaxy.

For the intrinsic samples, the filled circles show the values at their birth, while the open circles show the final stage ($z=0.17$).
Most importantly, all the intrinsic-associated objects (green circles) are DM-free from the birth to the final stage.
Together with the fact that they have high metallicities that are typical of massive galaxies, this supports the idea that they are not born as galaxies but as star clumps of a massive galaxy.

On the contrary, the intrinsic-isolated objects (blue circles) are born with a dark halo.
Some lose their DM halo while others leave the halo  through cosmic evolution.
In the end, the intrinsic-isolated objects are almost DM-free like their associated counterparts, but in essence, they are different.

\subsection{Formation pathways}
\label{subsection:distinguish_pathway}

The purpose of this study was to pin down the origin of compact stellar systems; most importantly, whether they were born intrinsically compact or stripped to become compact. 
NH covers a small volume yet shows a diversity of the formation pathways of compact systems, even down to the mass range of UCDs.
We have checked in the previous sections whether size, metallicity, age, and stellar/dynamical masses provide a hint for their origins. 

Two objects have mass compatible with the observed compact ellipticals, i.e., $M_*/\rm M_\odot \sim 10^{8-10}$.
In \NH, both of them are the remaining centers of much more massive dwarf galaxies after tidal stripping.

The origins of the other objects (53 UCD candidates) need a more careful inspection. 
The DM content or dynamical mass does not provide much information on their formation pathways, because both intrinsic and stripped systems are virtually devoid of DM when they are detected as compact systems.
The intrinsic UCD candidates show a bimodal metallicity distribution.
The metal-rich (32) intrinsic-associated objects have a metallicity typical of a massive galaxy, i.e., $[\rm{Z/H}] \sim 0$.
Their stellar ages are in the range of 1--7\,Gyr which is also typical of the stellar age of the disk of a massive galaxy.
The distance to their neighboring massive galaxy is $17^{\rm +61}_{\rm -9}\, \rm kpc$, which is well within one virial radius of a massive galaxy.
None of them has a meaningful dark halo associated. 
As a result, if we define a galaxy as a stellar system with a dark halo, these associated compact systems are not galaxies but star clumps belonging to a massive galaxy.

Seven of the intrinsic objects were formed without a massive neighboring galaxy, hence classified as ``intrinsic-isolated''.
They are born in their own dark halo.
Most of them have large ages (mean stellar age $\sim 7.8$ Gyr) and thus formed early, just like more massive galaxies. 
They all have a low metallicity (mean stellar metallicity $\sim-1.5$), which makes sense considering their birth positions.
The distance to their nearest massive galaxy is $d_{\rm host} = 160^{\rm +614}_{\rm -65}$\,kpc when they are detected finally as compact objects.
Many of them were later captured by the potential well of a massive galaxy, while some are still away from any significant dark halo.
At the final stage (i.e. when they are observed), most of them have no or negligible amount of dark matter within $3\,R_{\rm eff}$ because of the tidal stripping that was more effective to dark matter than to stars.
We conclude that the isolated intrinsically-compact objects are the DM-stripped result of dwarf galaxies born away from massive galaxies. 

Finally, we discuss the stripped compact objects. 
They appear somewhat distinct in the age-metallicity plane.
They are as old as isolated intrinsic compact systems but much more metal-rich. 
Their DM content (within 3\,$R_{\rm eff}$) has decreased by orders of magnitudes since their maximum-size epoch. 
They all are well within one virial radius of a typical massive galaxy, $d_{\rm host} = 84^{\rm +197}_{\rm -39}$\, kpc.
They seem compatible with the typical satellite dwarf galaxies near a massive galaxy.
After their early birth, they get their dark halo tidally stripped through their orbital motion inside the host galaxy halo.

For the two cE candidates, the median values of stellar ages in the final stage are $3.4_{\rm -0.5}^{+1.1}$ and $5.7_{\rm -1.0}^{+1.6}$\,Gyr (from 16\% through 84\% in the age distribution).
The large dispersions are compatible with the extended star-formation histories found in compact ellipticals \citep{Ferre-Mateu2018,Ferre-Mateu2021}.
For the UCD candidates, the age dispersion (16--84\%) is strikingly small, i.e., $2.3$\,Myr, which effectively indicates a single burst.
We find no clear difference between the stripped and the intrinsic samples in UCD candidates.
Regardless of their pathways, UCDs in \NH\ are single-burst objects.
In the case of stripped UCD candidates, the mean age dispersion at the time of their maximum size was 1.1\,Gyr, almost 3 orders of magnitudes larger than the value at the final stage.
This is because the stripped UCD candidates are the remnant star clumps of dwarf galaxies, rather than the stripped dwarf galaxies themselves.

\section{Summary}

Compact stellar systems such as ultra-compact dwarfs and compact ellipticals have properties in between ordinary galaxies and globular clusters, and their origin as to whether they are intrinsically compact or made compact later is a subject of heated debates.
We have used the recent \NH\ simulation to study the issue.
Thanks to its high spatial and mass resolutions, we can inspect low-mass objects down to the regime of UCDs. 
Besides, the high time cadence of the outputs of 15\,Myr allows us to monitor the motion and change of low-mass systems in great detail.
We specifically target the stellar systems smaller than the ordinary elliptical galaxies and with a stellar mass in the range of $M_{*} \sim 10^{6-9}\, {\rm M_{\odot}}$.
A careful inspection of the samples against numerical artifacts resulted in 55 compact stellar systems.
As a result, we have 53 UCD candidates $10^{6} \leq M_{*}/{\rm M_{\odot}} < 10^{8}$) and 2 cE candidates ($M_{*}/{\rm M_{\odot}} \geq 10^8$).
The main results can be summarized as follows.

All of the UCD candidates are a result of a single quick starburst, while the cE candidates show a substantially more extended star formation history.
Regarding their origins, we found both stripped and intrinsic cases.
The two cE candidates are born diffuse but later transformed to be compact via tidal stripping. 
Although our mass range is much smaller, \citet{Deeley2023} also found stripped cases of cE candidates in the TNG simulation.
Of the UCD candidates, 39 out of 53 are born compact nearly as they are today, hence ``intrinsically" compact.

If we divide the intrinsically compact systems into two sub-samples based on their birth environment, the systems born in isolated environments (\textbf{\em ``intrinsic-isolated''}: N=7) are older and metal-poor compared to those born inside or near a massive host galaxy (\textbf{\em ``intrinsic-associated''}: N=32).
Most of the intrinsically compact objects have a negligible amount of dark matter within $3 R_{\rm eff}$ in the final stage (when observed). 
The isolated objects are born as star clumps inside a dwarf galaxy with a dark halo, \textbf{hence qualifying as classic galaxies;} but the dark halos get stripped through tidal interactions with neighboring galaxies in due course.
The intrinsic compact objects born associated with a massive galaxy on the contrary do not have a dark halo throughout their entire life.
They are the surviving star clumps of their host galaxy, \textbf{rather than galaxies in the classical sense}.

The tidally stripped samples (\textbf{\em ``stripped''}: N=16) in the \NH\ simulation originate from the dwarf galaxies of $M_{*} \leq 10^{8-10}\ {\rm M_{\odot}}$ with a dark halo. 
Most of their mass in the envelope is stripped when they orbit around a massive galaxy over a timescale of 1 Gyr. 
Tidal stripping affects dark matter much more than star particles, and eventually, the dynamical mass of the final product becomes virtually the same as the stellar mass within $3 R_{\rm eff}$.

The gap between galaxies and star clusters is gradually being filled, which makes it even more confusing to determine what defines a ``galaxy'' than ever before.
Our study may raise a related question to this issue.
The stellar systems residing in the ``gap'' appear to form through a complex mixture of diverse formation pathways.

Our investigation is based on the \NH\ simulation which covers only a small volume of a field environment, whereas most of the observed data of compact objects are from cluster regions.
It is essential to perform a high-resolution simulation on dense regions to study the environmental effect on the formation processes of compact systems.

\begin{acknowledgments}
\section*{Acknowledgements}

We thank Sugata Kaviraj and Sebastien Peirani for constructive comments. This work was granted access to the HPC resources of CINES under the allocations  c2016047637, A0020407637 and A0070402192 by Genci, KSC-2017-G2-0003, KSC-2020-CRE-0055 and KSC-2020-CRE-0280 by KISTI, and as a “Grand Challenge” project granted by GENCI on the AMD Rome extension of the Joliot Curie supercomputer at TGCC.
The large data transfer was supported by KREONET which is managed and operated by KISTI. 
S.K.Y. acknowledges support from the Korean National Research Foundation (NRF-2020R1A2C3003769). 
S.C.R acknowledges support from the National Research Foundation of Korea (NRF-2022R1A2C1007721)
J.R. was supported by the KASI-Yonsei Postdoctoral Fellowship and was supported by the Korea Astronomy and Space Science Institute under the R\&D program (Project No. 2023-1-830-00), supervised by the Ministry of Science and ICT.
TK was supported by the National Research Foundation of Korea (NRF-2020R1C1C1007079). 
S.K. acknowledges support from the Korean National Research Foundation (NRF-2022R1C1C2005539).
This study was funded in part by the NRF-2022R1A6A1A03053472 grant and the BK21Plus program.

\vspace{-0.5cm}
\end{acknowledgments}


\bibliography{ref}{}
\bibliographystyle{aasjournal}


\begin{longtable*}[t]{llllllll}
\caption{General properties of the compact stellar systems}
\label{tab:Table1}\\

\hline
\hline
ID$^{a}$ & $log\ M_{\rm *}^{b}$ & $log\ R_{\rm eff}^{c}$ & $Group^{d}$ & $log\ M_{\rm *,max}^{e}$ & $log\ R_{\rm max}^{f}$ & $Age^{g}$ & $\rm {[Z/H]}^{h}$  \\ [5pt]
\hline
\hline
1 & 8.78 & 2.29 & stripped & 9.30 & 3.25 & $3.36^{+0.49}_{-1.05}$ & $-0.09^{+0.16}_{-0.16}$ \\ [3.5pt]
2 & 8.24 & 2.11 & stripped & 8.93 & 3.30 & $5.73^{+0.98}_{-1.57}$ & $-0.20^{+0.14}_{-0.13}$ \\ [3.5pt]
3 & 7.90 & 2.42 & isolated & - & - & $3.70^{+0.00}_{-0.00}$ & $-1.36^{+0.13}_{-0.04}$ \\ [3.5pt]
4 & 7.79 & 2.25 & isolated & - & - & $9.78^{+0.00}_{-0.00}$ & $-1.80^{+0.08}_{-0.06}$ \\ [3.5pt]
5 & 7.75 & 1.99 & stripped & 9.33 & 3.42 & $7.02^{+0.00}_{-0.00}$ & $-0.17^{+0.00}_{-0.00}$ \\ [3.5pt]
6 & 7.69 & 2.12 & associated & - & - & $7.44^{+0.00}_{-0.00}$ & $-0.65^{+0.01}_{-0.00}$ \\ [3.5pt]
7 & 7.59 & 1.98 & stripped & 8.33 & 3.18 & $8.01^{+0.00}_{-0.00}$ & $-0.65^{+0.01}_{-0.01}$ \\ [3.5pt]
8 & 7.60 & 1.80 & associated & - & - & $5.04^{+0.51}_{-0.33}$ & $-0.14^{+0.01}_{-0.02}$ \\ [3.5pt]
9 & 7.41 & 1.87 & associated & - & - & $2.42^{+0.00}_{-0.00}$ & $-0.57^{+0.01}_{-0.01}$ \\ [3.5pt]
10 & 7.36 & 2.01 & associated & - & - & $2.42^{+0.00}_{-0.00}$ & $-0.54^{+0.00}_{-0.00}$ \\ [3.5pt]
11 & 7.34 & 1.95 & associated & - & - & $4.59^{+0.00}_{-0.00}$ & $-0.14^{+0.00}_{-0.00}$ \\ [3.5pt]
12 & 7.30 & 1.82 & isolated & - & - & $7.10^{+0.00}_{-0.00}$ & $-1.82^{+0.00}_{-0.09}$ \\ [3.5pt]
13 & 7.27 & 1.68 & associated & - & - & $3.09^{+0.50}_{-0.00}$ & $-0.01^{+0.00}_{-0.11}$ \\ [3.5pt]
14 & 7.24 & 1.76 & associated & - & - & $1.33^{+0.00}_{-0.00}$ & $-0.22^{+0.01}_{-0.00}$ \\ [3.5pt]
15 & 7.21 & 1.86 & stripped & 8.74 & 3.50 & $9.46^{+0.60}_{-0.41}$ & $-0.83^{+0.44}_{-0.15}$ \\ [3.5pt]
16 & 7.18 & 1.67 & associated & - & - & $0.92^{+0.00}_{-0.00}$ & $-0.50^{+0.00}_{-0.00}$ \\ [3.5pt]
17 & 7.14 & 2.00 & associated & - & - & $1.56^{+0.00}_{-0.00}$ & $-0.34^{+0.00}_{-0.00}$ \\ [3.5pt]
18 & 7.14 & 2.02 & associated & - & - & $7.04^{+0.00}_{-0.00}$ & $-0.97^{+0.00}_{-0.02}$ \\ [3.5pt]
19 & 7.11 & 1.72 & associated & - & - & $6.77^{+0.00}_{-0.00}$ & $-0.38^{+0.00}_{-0.00}$ \\ [3.5pt]
20 & 7.12 & 1.92 & isolated & - & - & $9.84^{+0.46}_{-0.00}$ & $-2.55^{+0.00}_{-1.19}$ \\ [3.5pt]
21 & 7.07 & 1.81 & associated & - & - & $4.33^{+0.00}_{-0.00}$ & $-0.46^{+0.00}_{-0.00}$ \\ [3.5pt]
22 & 7.07 & 1.39 & associated & - & - & $1.17^{+0.00}_{-0.00}$ & $-0.16^{+0.00}_{-0.00}$ \\ [3.5pt]
23 & 7.06 & 1.89 & associated & - & - & $7.12^{+0.00}_{-0.00}$ & $-0.32^{+0.00}_{-0.00}$ \\ [3.5pt]
24 & 7.05 & 2.33 & stripped & 8.86 & 3.51 & $5.90^{+0.20}_{-0.00}$ & $-0.60^{+0.01}_{-0.09}$ \\ [3.5pt]
25 & 7.05 & 1.91 & stripped & 8.74 & 3.50 & $9.47^{+0.00}_{-0.00}$ & $-0.93^{+0.00}_{-0.02}$ \\ [3.5pt]
26 & 7.05 & 1.76 & associated & - & - & $5.88^{+0.00}_{-0.00}$ & $-0.34^{+0.01}_{-0.01}$ \\ [3.5pt]
27 & 7.05 & 1.89 & associated & - & - & $6.30^{+0.00}_{-0.00}$ & $-0.27^{+0.00}_{-0.00}$ \\ [3.5pt]
28 & 7.04 & 1.89 & associated & - & - & $7.24^{+0.00}_{-0.00}$ & $-0.84^{+0.00}_{-0.00}$ \\ [3.5pt]
29 & 7.03 & 1.70 & associated & - & - & $0.91^{+0.00}_{-0.00}$ & $-0.24^{+0.00}_{-0.00}$ \\ [3.5pt]
30 & 7.02 & 2.23 & isolated & - & - & $9.22^{+0.00}_{-0.00}$ & $-1.45^{+0.04}_{-0.05}$ \\ [3.5pt]
31 & 7.02 & 1.90 & associated & - & - & $5.56^{+0.00}_{-0.00}$ & $-0.25^{+0.00}_{-0.00}$ \\ [3.5pt]
32 & 7.00 & 2.20 & associated & - & - & $9.59^{+0.00}_{-0.00}$ & $-0.98^{+0.03}_{-0.04}$ \\ [3.5pt]
33 & 6.99 & 2.54 & stripped & 8.79 & 3.44 & $7.12^{+0.00}_{-0.00}$ & $-0.60^{+0.01}_{-0.01}$ \\ [3.5pt]
34 & 6.98 & 1.90 & stripped & 8.34 & 3.40 & $9.27^{+0.00}_{-0.00}$ & $-1.10^{+0.00}_{-0.00}$ \\ [3.5pt]
35 & 6.97 & 2.16 & stripped & 7.77 & 3.14 & $8.07^{+0.00}_{-0.00}$ & $-0.91^{+0.00}_{-0.00}$ \\ [3.5pt]
36 & 6.94 & 1.78 & stripped & 9.22 & 3.50 & $6.97^{+0.38}_{-0.00}$ & $-0.41^{+0.00}_{-0.05}$ \\ [3.5pt]
37 & 6.94 & 2.16 & isolated & - & - & $8.66^{+0.00}_{-0.00}$ & $-2.09^{+0.00}_{-0.01}$ \\ [3.5pt]
38 & 6.92 & 1.87 & associated & - & - & $6.83^{+0.00}_{-0.00}$ & $-0.18^{+0.01}_{-0.00}$ \\ [3.5pt]
39 & 6.89 & 1.78 & associated & - & - & $1.48^{+0.00}_{-0.00}$ & $-0.21^{+0.00}_{-0.00}$ \\ [3.5pt]
40 & 6.89 & 2.13 & associated & - & - & $6.98^{+0.00}_{-0.00}$ & $-0.33^{+0.00}_{-0.01}$ \\ [3.5pt]
41 & 6.86 & 1.26 & associated & - & - & $1.17^{+0.00}_{-0.00}$ & $-0.22^{+0.00}_{-0.00}$ \\ [3.5pt]
42 & 6.86 & 2.12 & stripped & 9.33 & 3.53 & $7.18^{+0.00}_{-0.00}$ & $-0.76^{+0.00}_{-0.00}$ \\ [3.5pt]
43 & 6.85 & 2.24 & stripped & 8.18 & 2.80 & $9.04^{+0.00}_{-0.00}$ & $-0.64^{+0.01}_{-0.01}$ \\ [3.5pt]
44 & 6.85 & 2.08 & stripped & 8.50 & 3.38 & $8.59^{+0.00}_{-0.00}$ & $-0.86^{+0.00}_{-0.03}$ \\ [3.5pt]
45 & 6.85 & 1.63 & associated & - & - & $1.15^{+0.00}_{-0.00}$ & $-0.13^{+0.00}_{-0.00}$ \\ [3.5pt]
46 & 6.83 & 2.29 & isolated & - & - & $6.87^{+0.00}_{-0.00}$ & $-1.20^{+0.02}_{-0.01}$ \\ [3.5pt]
47 & 6.83 & 2.12 & associated & - & - & $7.21^{+0.00}_{-0.00}$ & $-0.63^{+0.00}_{-0.00}$ \\ [3.5pt]
48 & 6.81 & 1.91 & associated & - & - & $2.32^{+0.00}_{-0.00}$ & $-0.18^{+0.00}_{-0.00}$ \\ [3.5pt]
49 & 6.80 & 1.91 & associated & - & - & $1.43^{+0.00}_{-0.00}$ & $-0.26^{+0.00}_{-0.00}$ \\ [3.5pt]
50 & 6.79 & 1.92 & associated & - & - & $2.81^{+0.00}_{-0.00}$ & $0.06^{+0.00}_{-0.00}$ \\ [3.5pt]
51 & 6.79 & 2.23 & stripped & 8.12 & 3.56 & $7.99^{+0.00}_{-0.00}$ & $-0.99^{+0.00}_{-0.00}$ \\ [3.5pt]
52 & 6.76 & 2.13 & associated & - & - & $5.54^{+0.00}_{-0.00}$ & $-0.21^{+0.00}_{-0.00}$ \\ [3.5pt]
53 & 6.75 & 1.85 & associated & - & - & $3.46^{+0.00}_{-0.00}$ & $-0.09^{+0.00}_{-0.00}$ \\ [3.5pt]
54 & 6.74 & 1.81 & associated & - & - & $4.11^{+0.00}_{-0.00}$ & $-0.04^{+0.00}_{-0.00}$ \\ [3.5pt]
55 & 6.72 & 2.46 & stripped & 9.15 & 3.53 & $2.57^{+0.00}_{-0.00}$ & $-0.34^{+0.00}_{-0.01}$ \\ [3.5pt]
\hline 
\hline 
\end{longtable*}

\raggedright
\tablenotetext{a}{The identification number of the compact stellar systems.}
\tablenotetext{b}{The stellar mass inside the half-mass radius in $\rm M_\odot$.}
\tablenotetext{c}{The half-mass radius of the stellar system in pc.}
\tablenotetext{d}{The grouping of each stellar system.}
\tablenotetext{e}{The stellar mass inside the half-mass radius at the maximum epoch in $\rm M_\odot$ (for stripped sample).}
\tablenotetext{f}{The half-mass radius of the stellar system at the maximum epoch in pc (for stripped sample).}
\tablenotetext{g}{The median age and the 16 and 84\% percentiles of the stellar particles inside 3$R_{\rm eff}$ in Gyr.}
\tablenotetext{h}{The median metallicity and the 16 and 84\% percentiles of the stellar particles inside 3$R_{\rm eff}$.}


\end{document}